\def\fdeg{\hbox{$.\!\!^{\circ}$}}

\documentstyle[12pt,aasms4]{article}

\begin{document}

\title{Interstellar Gas\\in\\ 
Low Mass Virgo Cluster Spiral 
Galaxies\footnote{Based
on observations made with ISO, an ESA project with
instruments funded by ESA Member States and with
the participation of ISAS and NASA.}}
\author{Beverly J. Smith}
\affil{IPAC/Caltech, MS 100-22, Pasadena CA  91125}
\centerline{and}
\author{Suzanne C. Madden}
\affil{Centre d'Etudes de Saclay, Service d'Astrophysique,
Ormes des Merisiers, 91191 Gif-Sur-Yvette Cedex, France}
\vskip 1.0in

\begin{abstract}

We have measured the
strengths of the [C~II] 158 $\mu$m,  [N~II] 122 $\mu$m,
and CO~(1~$-$~0) lines  
from five low blue luminosity spiral galaxies in
the Virgo Cluster,
using the Infrared Space Observatory and
the NRAO 12m millimeter telescope.
Two of the five galaxies have
high 
L([C~II])/L(CO) 
and 
L(FIR)/L(CO)
ratios compared to higher mass spirals. 
These two galaxies, NGC 4294 and NGC 4299, 
have
L([C~II])/L(CO) 
ratios of $\ge$14,300 and 15,600,
respectively, which are similar to values
found in dwarf irregular galaxies.
This is the first time that
such enhanced L([C~II])/L(CO) ratios have been found in spiral galaxies.
This result
may be due to low abundances of dust and heavy elements,
which can cause
the CO (1 $-$ 0) 
measurements to underestimate the molecular gas content.
Another possibility is that radiation
from diffuse HI clouds may dominate 
the 
[C~II] 
emission 
from these galaxies.
Less than a third of the observed [C~II] emission
arises from HII regions.

\end{abstract}


\section{Introduction}

The 2.6mm CO (1 - 0) line is commonly used as a measure of the mass
of interstellar molecular hydrogen in external galaxies (e.g., 
\markcite{y90}Young
1990; \markcite{ys91}Young 
$\&$ Scoville 1991).  In 
these studies, the CO intensity is generally assumed to
be proportional to the H$_2$ column density, with a constant of
proportionality equal to that found for Galactic clouds. Recent
theoretical and observational evidence, however, suggests that this
ratio may vary significantly from galaxy to galaxy and within galaxies,
as a function of the ambient radiation field, the metal abundance,
and the dust extinction.  In 
particular, low abundances of C and O
and a low dust column density may 
result in deeper penetration of ultraviolet 
photons, decreasing the 
size of the CO-emitting region
within a molecular cloud, while
H$_2$ remains relatively self-shielded.
This can result in large volumes of
molecular clouds which are not
sampled by CO observations
\markcite{mb88}(Maloney
$\&$ Black 1988; 
\markcite{m90}Maloney 1990; 
\markcite{mw96}Maloney $\&$ Wolfire 1996). 
In dwarf irregular galaxies, the CO intensity
is generally faint relative to the star formation rate 
\markcite{c86}(Combes 1986;
\markcite{ty87}Tacconi 
$\&$ Young 1987), perhaps due to this 
effect rather than to a particularly low
H$_2$ content.  CO studies
of individual clouds and cloud complexes in dwarf irregular galaxies are
consistent with this scenario; the virial masses implied by the
line widths are often higher than the H$_2$ masses indicated by the
CO luminosities and the standard 
I$_{CO}$/N$_{H_2}$ conversion 
factor
(e.g., \markcite{dh89}Dettmar $\&$ Heithausen 1989; 
\markcite{r91}Rubio et al. 1991; \markcite{wr92}Wilson $\&$ 
Reid 1992). These investigations suggest that 
I$_{CO}$/N$_{H_2}$
is correlated with metallicity, but with significant scatter 
\markcite{o93}(Ohta
et al. 1993).

Other evidence for a lower 
I$_{CO}$/N$_{H_2}$ conversion 
factor
in irregular galaxies compared to the Milky Way is the detection of strong
[C~II] 158 $\mu$m emission relative to the CO (1~-~0) intensity
\markcite{m94}(Mochizuki et al. 1994; 
\markcite{p95}Poglitsch et al. 1995; 
\markcite{i96}Israel et al. 1996; 
\markcite{m97}Madden et al. 1997), compared to the
L([C~II])/L(CO) ratio for Galactic star forming regions 
and the central regions of high mass spiral galaxies
\markcite{c85}(Crawford et al. 1985;
\markcite{s91}Stacey et al. 1991).
This C$^+$ line is an important interstellar cooling
line, and in regions of 
on-going star formation it originates
mainly from warm (T $\sim$ 300K) dense (10$^2$ $-$ 10$^4$ cm$^{-3}$)
gas in photodissociation regions (PDRs)
which form the boundaries between HII regions
and molecular clouds 
\markcite{s91}(Stacey et al. 1991).
In the inner regions of high mass galaxies, 
the observed L([C~II])/L(CO) and
L([C~II])/L(FIR) ratios 
\markcite{s91}(Stacey et al. 1991)
agree
with those
predicted by
theoretical studies 
of PDRs 
in gas with solar abundances
\markcite{th85}(Tielens $\&$ Hollenbach 1985;
\markcite{w89}Wolfire,
Hollenbach, $\&$ Tielens 1989).
For
dwarf irregular galaxies, however, 
the L([C~II])/L(CO) ratios
do not fit these models
\markcite{m94}(Mochizuki et al. 1994; 
\markcite{p95}Poglitsch et al. 1995; 
\markcite{i96}Israel et al. 1996; 
\markcite{m97}Madden et al. 1997), 
probably because of the low metallicities of these systems.
In gas with low
abundances and dust content, UV radiation penetrates more
deeply into a molecular cloud, causing a larger C$^+$ region relative to
the CO core,  
and therefore elevated
global L([C~II])/L(CO) ratios 
\markcite{m90}(Maloney 1990; \markcite{mw96}Maloney
$\&$ Wolfire 1996). 
It is in this C$^+$
emitting region that H$_2$ may be present
and unaccounted for by CO observations.

If decreased I$_{CO}$/N$_{H_2}$
ratios are common in low metallicity
systems, then one would expect to find them not only in
dwarf irregulars, but also in low mass spiral galaxies, since
metallicity appears to be correlated with galaxian mass 
\markcite{pe81}(Pagel
$\&$ Edmund 1981; 
\markcite{gs87}Garnett $\&$ Shields 1987; 
\markcite{ve92}Vila-Costas $\&$ Edmunds
1992). A good sample to search for this effect is the low mass
Virgo Cluster spirals studied by 
\markcite{ky88}Kenney $\&$ Young (1988).  These
galaxies have high star formation rates, as measured by their
far-infrared and H$\alpha$ luminosities, relative to their CO
luminosities, compared to high mass Virgo galaxies. 
\markcite{ky88}Kenney $\&$ Young
(1988) suggest that 
the total gas mass,
rather than the molecular gas alone, is important in determining the
rate of star formation.
An alternative explanation is that the amount
of molecular gas is
underestimated in the low mass spirals by the CO~(1~$-$~0)
line 
\markcite{ky88}(Kenney $\&$ Young 1988, \markcite{ky89}1989).

To investigate the physical conditions of the interstellar
medium in these galaxies
and to compare with dwarf irregulars and high mass spirals,
we have used the Long Wavelength Spectrometer
\markcite{c96}(LWS) (Clegg et al. 1996)
on the Infrared Space Observatory (ISO)
\markcite{k96}(Kessler et al. 1996)
to measure the [C~II]
158 $\mu$m emission from five of the low mass  
\markcite{ky88}Kenney
$\&$ Young (1988) Virgo spiral galaxies.
To distinguish between [C~II] radiation
from neutral and ionized gas, we have also observed the [N~II]
122~$\mu$m line, which originates only in HII regions.
Several of these galaxies were undetected in CO (1~$-$~0)
in the 
\markcite{ky88}Kenney $\&$ Young (1988) survey.  
We have therefore also used the National Radio Astronomy
Observatory 
(NRAO\footnote{The National Radio Astronomy Observatory
is operated by Associated Universities, Inc., under
cooperative agreement with the National Science Foundation.})
12m 
telescope to make more sensitive measurements of
this line.
These results are compared
with the IRAS far-infrared fluxes and published
H$\alpha$ and 
HI fluxes, along with results for other galaxies and theoretical
models of C$^{+}$ emission.

The sample galaxies are listed in Table 1, along
with their positions, types, total blue magnitudes,
sizes, and velocities.
In Table 2, we provide total far-infrared flux densities
from IRAS along with
far-infrared luminosities.
Table 2 also contains 
blue and H$\alpha$ luminosities and HI masses.

\section{Observations and Data Reduction}

\subsection{Infrared Space Observatory}

On 1996 July 12,
the ISO LWS observed
the [C II] 158 $\mu$m  
line in
the five sample galaxies.
The 
[N II] 122 $\mu$m line was also observed
in three of the galaxies.
Details on these observations are given
in Table 3.
The observed positions are given in Table 1.
To measure foreground Galactic emission at these locations,
we also observed nearby (10$'$ away) sky locations 
with the
same observing parameters.
These off-galaxy observations were concatenated
onto the galaxy observations in a single
observing sequence.  The neighboring 
galaxies NGC 4189 and NGC 4222 shared a single
sky measurement and were observed together in
one concatenation chain, as were NGC 4294 and NGC 4299.
We used
the medium resolution grating mode, with a spectral
scan width of 2 (2 spectral elements on either side of
the line), giving a total of 5 resolution elements per scan,
and a sampling interval of 4.  The
spectral resolution with this setup is 0.6 $\mu$m,
the total observed bandpass is 2.8 $\mu$m, and the sampling
interval is 0.14 $\mu$m.
At this resolution (1500 km s$^{-1}$), the lines
are expected to be unresolved.  
The `fast' scanning mode was used for these observations.
The ISO LWS beam is
FWHM
$\sim$80$''$ 
\markcite{s96}(Swinyard et al. 1996).

These data were processed by version 6.0 of the ISO pipeline,
and were then further 
reduced (deleting
bad data,
summing scans) at the Infrared Processing
and Analysis Center (IPAC) using version 
1.2a
of the Interactive
Spectral Analysis Package (ISAP).
The LWS
version 6 pipeline includes a calibration correction
based on internal illuminator flashes and a `drift' correction
for changing detector responsivities based on the source
intensities.
The absolute calibration uncertainties are expected
to be less than 30$\%$.

\subsection{NRAO 12m Millimeter Telescope}
	
The sample galaxies were observed in the CO (1 $-$ 0) line
on 1996 April 10-13,
May 9-10, and December 14 and 20
with the 3 mm SIS receiver
on the NRAO 12m telescope.
At this frequency, the beamsize FWHM of the telescope is 55$''$.
Two 256 $\times$ 2 MHz filterbanks, one for each polarization, were used
for these observations,
providing a total bandpass of 1330 km s$^{-1}$ and a spectral resolution
of 5.2 km s$^{-1}$.
The central velocity used for each galaxy is
given in Table 1.

The positions observed with the 12m telescope
are given in Table 5.
For the two brightest galaxies, NGC 4189
and NGC 4522, we made a 5 point map 
with 25$''$ spacing.
For the fainter galaxies, we only observed the central point.
Calibration was accomplished using an ambient chopper wheel.
The observations were made using a nutating subreflector with a beam
throw of
3$'$ in the azimuthal direction.
The pointing was checked periodically with 
bright continuum sources,
and 
was consistent to $\sim$10$''$.
The CO spectra for each position were summed and a linear
baseline subtracted.

\section{Results}

\subsection{The ISO Data}

The [C~II] 158 $\mu$m spectra for
the five galaxies 
are given in Figure 1 and the measured
line fluxes are given in Table 4.  All five sample
galaxies were strongly detected in this line.
The off positions show no line emission and significantly
fainter continuum emission (1 $-$ 4 $\times$
10$^{-20}$ W cm$^{-2}$ $\mu$m$^{-1}$) than the galaxies.
The [N~II] 122 $\mu$m data for
the three galaxies 
observed in this line
are presented in Figure 2 and Table 4.
Only NGC 4189 was detected in [N~II].
As with [C~II], the off positions show no
[N~II] emission and much weaker 
continuum emission 
(6 $-$ 12 times fainter)
than the galaxies.

\subsection{CO (1 $-$ 0) Data}

The final CO spectra 
for the central positions in the galaxies
are shown in Figure 1.
In Table 5, we list the rms noise level T$_R$$^*$, the
integrated
CO flux densities $I_{CO} = \int{T_R^*dV}$, 
the peak temperature, and the 
FWZM line width for all of the observed positions.
We detected four out of the five galaxies.
In the 
\markcite{ky88}Kenney $\&$ Young (1988) CO (1 $-$ 0) study of these
galaxies, only NGC 4189 was detected.  When converted
to the same units, our central flux for
this galaxy agrees with their value
as tabulated in
\markcite{y95}Young et al. (1995).

\section{Discussion}

\subsection{Comparison with Other Galaxies}

In Table 6, we list some derived parameters for these galaxies,
including M(H$_2$), assuming the standard Galactic
I$_{CO}$/N$_{H_2}$ ratio 
\markcite{b86}(Bloemen et al. 1986), L([C~II]),
L([N~II]), 
L(FIR)/M(H$_2$),
and L([C~II])/L(CO).  
We also give a lower limit to the mass of hydrogen
gas associated with the observed C$^+$ emission, 
M$_{C^+}$$^{min}$(H) = 0.45 L([C~II]), where
M$_{C^+}$$^{min}$(H) and L([C~II]) are measured in
solar units.
This relationship is derived in the high density,
high temperature limit (n $>>$ 3000 cm$^{-3}$; T $>>$ 91K)
as in
\markcite{c85}Crawford et al. (1985), assuming solar
carbon abundance and that all the C is in the form of C$^+$ ([C$^+$]/[H]
= 3 $\times$ 10$^{-4}$).
For lower temperatures, densities, and abundances, higher masses are
expected.

The [C~II] luminosities for these galaxies range from
5.7 $\times$ 10$^6$ L$_{\sun}$ to
1.1 $\times$ 10$^7$~L$_{\sun}$.
Inspection of the IRAS HiRes images of these
galaxies indicate that they are essentially
unresolved in the 85$''$ $\times$ 50$''$
60 $\mu$m HiRes beam, so
nearly all the far-infrared flux originates within the ISO beam.
The global L([C~II])/L(FIR)
ratios for these galaxies therefore range from 0.004 $-$ 0.008.
For comparison, the galaxies surveyed
by 
\markcite{s91}Stacey et al. (1991) 
have
L([C~II])/L(FIR) ratios between 0.0015 $-$ 0.006, while
in the Galactic plane
L([C~II])/L(FIR) $\sim$ 0.006 
\markcite{n95}(Nakagawa et al. 1995).
In the LMC
L([C~II])/L(FIR) is slightly higher, $\sim$ 1$\%$ 
\markcite{m94}(Mochizuki et al. 1994; \markcite{i96}Israel et al.
1996).
Global flux ratios for the spiral
galaxies NGC 6946 
\markcite{m93}(Madden et al.
1993) and NGC 5713 
\markcite{l96}(Lord et al. 1996)
are $\sim$0.011 and 0.007, respectively.
Thus the
L([C~II])/L(FIR)
ratios in the surveyed Virgo galaxies are not unusual
compared to other galaxies.

In contrast to 
L([C~II])/L(FIR),
the 
L([C~II])/L(CO) ratios for the sample galaxies
vary widely, ranging from 1650
in NGC 4189 to 15,600 in NGC 4299.
This variation is evident in Figure 1, where the
CO and [C~II] spectra are plotted together.
These differences are too large to be accounted for
by the calibration uncertainties alone.
The highest values of
L([C~II])/L(CO) 
are typical of the global
ratio found for the LMC, $\sim$23,000
\markcite{m94}(Mochizuki et al. 1994),
but are not as high as 
at selected HII regions within the
LMC, for example, 
at 30 Dor, where the ratio is $\sim$70,000
\markcite{p95}(Poglitsch et al. 1995).
The
L([C~II])/L(CO) ratios for NGC 4294 and NGC 4299 are 
higher than those typically found
in the central regions of high
mass galaxies, $\sim$4000 in starbursts 
and $\sim$1300 in quiescent galaxies
\markcite{s91}(Stacey et al. 1991).
In
the disk of the Milky Way, 
L([C~II])/L(CO) 
$\sim$ 1300 
\markcite{n95}(Nakagawa et al. 1995).

The observed
L([N II]) 122 $\mu$m/L([C II]) 
ratios 
are 0.078 
for NGC 4189 
and $\le$0.058 and $\le$0.069
for NGC 4294 and NGC 4299, respectively.
For comparison, this 
ratio is 0.2
in the starburst galaxy M82 (Petuchowski et al. 1994),
0.08 $-$ 0.1
in the Milky Way
\markcite{b94}(Bennett et al. 1994; 
\markcite{b96}Bennett 
1996),
$\le$0.18
in the merging pair NGC 4038/9
\markcite{f96}(Fischer et al. 1996),
and $\le$0.12 
in the spiral galaxy NGC 5713
\markcite{l96}(Lord et al. 1996).
Thus the high 
L([C~II])/L(CO)
galaxies appear somewhat weak
in [N~II] compared to more massive 
galaxies,
although
the paucity of available extragalactic [N~II] detections 
and the ISO calibration errors makes this
result uncertain.

In the two galaxies with the highest
L([C~II])/L(CO) ratios, NGC 4294 and NGC 4299,
the L(FIR)/M(H$_2$) ratios are 
more than 8 times larger than the mean
value found for the higher mass Virgo galaxies in 
\markcite{ky88}Kenney $\&$
Young (1988).
Thus 
two out of the five galaxies in this sample 
appear to be extreme in comparison to higher
mass galaxies, with significantly higher 
L([C~II])/L(CO)
and L(FIR)/M(H$_2$) 
values but low 
L([N~II]) 122 $\mu$m/L([C~II]) ratios.
L(B)/L(CO) is also enhanced in NGC 4294 and
NGC 4299 relative to the other galaxies
in the sample.
In contrast, 
the L([C~II])/L(FIR) ratio is only marginally
higher in NGC 4294 and NGC 4299.

\subsection{PDRs and Metallicity}

In the central regions of high mass spiral galaxies,
the L([C~II])/L(FIR)
and L([C~II])/L(CO) ratios are consistent
with theoretical models of emission
from PDRs
with solar abundances
\markcite{c85}(Crawford et al. 1985;
\markcite{s91}Stacey et al. 1991).
As noted previously, three of the galaxies
in our Virgo sample, NGC 4189, NGC 4222,
and NGC 4522, have 
L([C~II])/L(FIR)
and L([C~II])/L(CO) ratios similar to
those seen in high mass galaxies, indicating
that solar metallicity PDRs can also account for the [C~II]
emission from these galaxies.

Two of our galaxies, however, have extreme
L([C~II])/L(CO) values, more similar to
those found for low metallicity dwarfs than
for high mass spirals.
In general, their CO fluxes
are very weak relative to their far-infrared,
blue, and [C~II] fluxes,
suggesting that the difference between
these galaxies and the other three galaxies 
may be due to a deficiency in CO rather than
an excess in [C~II].

The similarity with dwarf galaxies suggests
that 
low metallicity and therefore
a low I$_{CO}$/N$_{H_2}$ ratio may be responsible.
In dwarf galaxies, enhanced
L([C~II])/L(CO) ratios are explained by low dust
and metal
content
\markcite{m94}(Mochizuki et al. 1994; 
\markcite{p95}Poglitsch et al. 1995; 
\markcite{i96}Israel et al. 1996; 
\markcite{m97}Madden et al. 1997).
Although no measurements of metallicity
are available for these Virgo
spirals at present, they
have relatively low blue luminosities which
suggests they may be metal-poor.
Their absolute blue magnitudes, $\sim$$-$18 to
$-$19, are similar to that of the LMC
and other dwarf galaxies which are known
to be metal-poor 
\markcite{skh89}(Skillman, Kennicutt, 
$\&$ Hodge 1989).
Less than solar metallicities
have also been seen in spiral and starburst galaxies
at these magnitudes 
\markcite{ve92}(Vila-Costas $\&$ Edmunds 1992;
\markcite{sck94}Storchi-Bergmann, Calzetti,
$\&$ Kinney 1994).

If metallicity is responsible, though, it is
surprising that 
NGC 4294 and NGC 4299 are weak in CO while 
the other three galaxies are not, since
all five galaxies have approximately the same
apparent blue magnitudes (Table 1).
The far-infrared flux densities are also similar (Table 2).
Variations in 
the 
L([C~II])/L(CO)
ratio could be caused by
the smaller CO beam
missing emission contained in the larger ISO beam,
however,
H$\alpha$ images 
of these galaxies
\markcite{hk83}(Hodge $\&$ Kennicutt 1983)
show that most of the star formation is occuring within
the inner arcminute,
suggesting
that most of the CO in these galaxies is contained in the 
central beam.

A possible explanation for the more extreme
L([C~II)/L(CO) values for NGC 4294 and NGC 4299 is
that they may be closer, and therefore have lower blue
luminosities and masses than the other three galaxies.
The Virgo Cluster has long been known to have a complicated
morphology (c.f., \markcite{bts87}Binggeli, Tammann, $\&$ Sandage 1987;
\markcite{pt88}Pierce $\&$ Tully 1988), and may have
considerable extent along the line of sight.
A recent Tully-Fisher study 
derives distances for individual galaxies in the Virgo Cluster
ranging from 12 to 30 Mpc
(\markcite{yfo97}Yasuda, Fukugita, $\&$ Okamura 1997).
For NGC 4294, they quote a distance of only 13.6 $\pm$ 1.1 Mpc,
while NGC 4189, NGC 4222, and NGC 4522 were found to be more distant,
at 33.8 $\pm$ 5.1 Mpc, 22.9 $\pm$ 1.5 Mpc, and 16.3 $\pm$ 1.1 Mpc,
respectively.  No Tully-Fisher distance was derived
for the face-on galaxy NGC 4299, however,
it may be in a bound pair with
NGC 4294, which is only a few radii away on the sky.
The optical and HI morphologies and HI kinematics
of this pair are disturbed, suggesting a recent
interaction 
\markcite{sbt85}(Sandage, Binggeli,
$\&$ Tammann 1985; \markcite{w88a}Warmels 1988a).
If NGC 4294 and NGC 4299 have lower blue luminosities
and masses than
the other three galaxies
then they may have lower metallicities as well, which
could explain the enhanced L([C~II])/L(CO) values.

There may
also be a spread in metallicity
among these galaxies, even if they have similar blue luminosities.
In the 
\markcite{vp90}van den Bergh
$\&$ Pierce (1990) 
study of dust in Virgo cluster
galaxies, 
NGC 4189, NGC 4222, and NGC 4522
are all classed as dusty or very dusty.  They
do not fit the average magnitude-dustiness relationship
for Virgo galaxies, being dustier than expected
\markcite{vp90}(van den Bergh
$\&$ Pierce 1990). 
This suggests that they may
be more metal-rich than typical for their magnitudes.  
Alternatively, an interaction
between NGC 4294 and NGC 4299 may have driven unprocessed
gas into the centers of these galaxies, causing
the inner regions
of these galaxies
to have significantly lower metallicities
than other galaxies of similar mass.
This question of metallicity 
should be tested with followup optical spectroscopy.

Another possible explanation for
the differences between the galaxies is that they
are in different evolutionary
stages.
Perhaps NGC 4294 and NGC 4299 are in a post-burst
stage where the molecular gas has been depleted but
PDR [C~II] emission is still bright.
Within the Large Magellanic Cloud, the
wide variation in the
observed
L([C~II])/L(CO) ratio has been
attributed to evolutionary differences
\markcite{i96}(Israel et al. 1996).
NGC 4294 and NGC 4299 have somewhat higher H$\alpha$
fluxes than the other three
galaxies (Table 2),
and therefore more star formation activity, if they are
all
at the same distance.
Their H$\alpha$ equivalent widths
are among the highest in the 
\markcite{kk83}Kennicutt $\&$ Kent (1983) galaxy sample,
similar to those of starburst galaxies, thus
they have a high current to past star formation rate.
The optical photographs of NGC 4294 and NGC 4299 presented in 
\markcite{sbt85}Sandage et al. (1985)
show disturbed spiral structure
containing numerous bright HII regions.
NGC 4294 and NGC 4299 also have higher 60 $\mu$m to 100 $\mu$m
flux ratios than the other three galaxies (Table 2), indicating warmer dust.
Thus strong ultraviolet radiation fields may
contribute to the enhanced L([C~II])/L(CO) ratios in
these galaxies.  However,
the observed ratios
are much
more extreme than those typically seen in starburst
galaxies \markcite{s91}(Stacey et al. 1991), 
indicating that 
star formation and gas depletion
are probably
not solely responsible for the observed line ratios.

\subsection{Emission from Diffuse HI Clouds}

An alternative explanation for the enhanced
L([C~II])/L(CO) ratios in NGC 4294 and NGC 4299
is that there is
a large component of C$^{+}$ associated with diffuse atomic
gas.
Spatially resolved [C~II] observations
of the nearby spiral galaxy NGC 6946
\markcite{m93}(Madden et al.
1993) revealed a very extended component containing
$\sim$70$\%$ of the total [C~II] emission of the galaxy.
This diffuse emission is 
thought to be due to C$^+$ arising from diffuse HI clouds
rather than standard PDRs.
If emission from diffuse atomic gas rather
than PDRs dominates
in these galaxies, then one would not expect
a strong correlation between CO and [C~II] fluxes.

At Virgo, the ISO beam covers an area 3.9 kpc in radius,
assuming a distance of 20 Mpc.
Extended [C~II] radiation is seen at this radius in
NGC 6946 \markcite{m93}(Madden et al. 1993).
Optical images 
of the Virgo galaxies
\markcite{sbt85}(Sandage et al. 1985)
show that the ISO beam covers not
just the nuclei but also most of the observed spiral arms,
so may contain considerable diffuse disk emission.
The spread in 
L([C~II])/M(HI) for the sample galaxies is only 
a factor of 3, compared to
a factor of 20 in L([C~II])/L(CO).
Thus the observed [C~II] fluxes are more closely correlated with
the total HI mass than with CO.

To compare the observed [C~II] and HI fluxes for these galaxies
more accurately,
it is first necessary to take into account the fact
that the observed HI distributions
extend beyond the ISO beam.
Four out of the 5 galaxies in our sample
have published interferometric HI maps 
\markcite{w88a}(Warmels 1988a; 
\markcite{c90}Cayette et al. 1990) in which 
HI is detected out to radii of $\sim$2$'$.
Over the ISO beam, the column density of HI is $\sim$10$^{21}$
cm$^{-2}$ in NGC 4189, NGC 4294, and NGC 4299,
and a factor of $\sim$3 times lower in NGC 4222
\markcite{w88b}(Warmels 1988b).
The fraction of the total
HI contained within an $\sim$80$''$$-$90$''$
diameter in these galaxies is 35$\%$ $\sim$ 64$\%$
\markcite{w88b}(Warmels 1988b), so
4 $-$ 8 $\times$ 10$^8$ M$_{\sun}$
of atomic hydrogen is contained within the ISO beam.
Thus our galaxies have 
M(HI)/L([C~II]) = 40 $-$ 90 M$_{\sun}$/L$_{\sun}$
within the ISO beam.
For comparison, the extended [C~II] emission
in NGC 6946 has  
M(HI)/L([C~II]) $\sim$ 16 M$_{\sun}$/L$_{\sun}$
(Madden et al. 1993).
Thus the amount of HI gas relative to the [C~II] flux
is even
higher in our sample galaxies
than in the outer disk of NGC 6946.

These galaxies are relatively rich in HI.
In NGC 4294 and NGC 4299 in particular
the interstellar
medium may be predominantly atomic if
CO is not greatly underestimating the molecular gas;
the 
Galactic I$_{CO}$/N$_{H_2}$ ratio gives M(HI)/M(H$_2$) 
ratios of $\ge$20 for these galaxies.
We note that in spite of the fact that these
galaxies are in the Virgo Cluster they
are not
extremely HI-deficient for
their types;
their HI contents, defined according
to the scale in
\markcite{gh83}Giovanelli $\&$ Haynes (1983),
are fairly typical of field galaxies
\markcite{ky88}(Kenney $\&$ Young 1988).
Only NGC 4522
appears somewhat HI deficient
\markcite{ky88}(Kenney $\&$ Young 1988).
The five sample
galaxies lie in the outer regions of the Virgo Cluster,
at cluster radii of 2\fdeg5 $-$ 4\fdeg5, and so are
not subject to as strong HI removal mechanisms
as galaxies in the cluster core.

As in the Milky Way
\markcite{kh87}(Kulkarni $\&$ Heiles 1987),
the atomic gas in external galaxies may exist in
two forms: cold HI clouds (cold neutral
medium; $\sim$80K) and warm diffuse 
HI (warm neutral medium; $\sim$6000K).
To estimate the amount of [C~II] arising from cold atomic
clouds in our sample galaxies,
we use the relationship M(HI)/L([C~II]) 
= [580/($n_H$e$^{-91/T}$)][3~$\times$~10$^{-4}$/X$_{C^+}$], where the
mass of HI and luminosity of [C~II] are in solar units,
n$_H$ is the density of H atoms in cm$^{-3}$, T is the kinetic
temperature of the gas in degrees K, and X$_{C^+}$ is
the abundance
[C$^+$]/[H].
This formula was derived from the relationship for collisional
excitation of the C$^+$ line in the optically thin
case 
\markcite{c85}(Crawford et al. 1985), in the low
density (n $<<$ 4 $\times$ 10$^3$ cm$^{-3}$) limit.
For cold HI clouds, collisional excitation by
electrons should not be important 
\markcite{c85}(Crawford et al. 1985;
\markcite{m93}Madden et al. 1993).
Assuming a solar abundance [C$^+$]/[H] = 3 $\times$ 10$^{-4}$
and 
typical values of T and n$_H$ for the cold neutral medium
of 80K and 90 cm$^{-3}$ \markcite{kh87}(Kulkarni
$\&$ Heiles 1987;
\markcite{m93}Madden et al. 1993) gives
an expected ratio of M(HI)/L([C~II]) 
= 20 M$_{\sun}$/L$_{\sun}$.

We therefore conclude that, if CNM 
with T $\sim$ 80K and n$_H$ $\sim$ 90 cm$^{-3}$ 
makes up $\sim$50$\%$ of the observed HI in these galaxies,
there are
sufficient cold atomic clouds in these galaxies to account
for all of the observed C$^+$ emission.  If the
atomic gas is colder and more diffuse, however, or if the
fraction of HI emission originating from the CNM is
lower, then 
only a fraction of the C$^+$ can
be accounted for by the CNM.
Likewise, 
if the metallicity is low, 
all of the [C~II] emission cannot
arise from the CNM.

The WNM, in contrast to the CNM, 
does not appear to be a 
major contributor to the observed [C~II] flux from
our galaxies.
Unlike in the CNM, in the WNM C$^+$ excitation by electrons is 
an important factor 
\markcite{m93}(Madden et al. 1993).
Including this term
as in equation 5 of \markcite{m93}Madden et al. (1993),
and including the critical density terms for collisions
with HI and electrons as in 
\markcite{lr77}Launay $\&$ Roeuff (1977)
and 
\markcite{hn84}Hayes $\&$ Nussbaumer (1984),
for the WNM the relationship between 
temperature, hydrogen density,
and 
HI and [C~II] 
emission is:
L([C~II])/M(HI)
= [3~$\times$~10$^{-4}$/X$_{C^+}$] $[ 2/(1 + 710/n_H) + 
2/(1 + 0.13T^{1/2}/({X}_{e}n_H))$ ],
where M(HI) and L([C~II])
are in solar units,
T is in degrees K, n$_H$ is in 
cm$^{-3}$, $X$$_e$ is the fractional ionization of hydrogen,
and X$_{C^+}$
is the C$^+$ abundance.
Assuming solar abundance and the 
Galactic WNM parameters of $X$$_e$ = 0.03, T = 6000K,
and n $\sim$ 1 cm$^{-3}$
\markcite{kh87}(Kulkarni $\&$ Heiles 1987)
gives 
a predicted relationship between HI mass and [C~II] luminosity
of $M(HI)/L([C~II]) \sim 120~M_{\sun}/L_{\sun}$.
This is higher than the observed ratios
in the Virgo galaxies, indicating
that there is not sufficient WNM in these galaxies
to account for the observed [C~II] emission.
This same conclusion was reached for the extended
emission in the more massive
spiral galaxy NGC 6946
\markcite{m93}(Madden et al. 1993).

\subsection{Ionized Gas}

[C~II] emission may also originate from HII regions, both
compact and diffuse.
HII regions with densities from 10$^2$ 
$-$ 10$^5$ cm$^{-3}$ have been
modeled in detail by \markcite{r85}Rubin (1985).
These models provide
expected L(H$\alpha$)/L([C~II])
ratios
for HII regions ionized by stars
with temperatures ranging from 31,000K $-$ 45,000K.
For solar abundance and 
a Lyman continuum photon rate of 10$^{49}$ photons s$^{-1}$,
this ratio ranges from 9.8 $-$ 16.1 for a 
density of 100 cm$^{-3}$ and a central star
temperature between 31,000K and 37,000K.
This ratio increases rapidly 
with density and stellar
temperature,
to 1.4 $\times$ 10$^5$
for a density of 10$^5$ cm$^{-3}$ and 
a stellar temperature of 40,000K, because
of an increased fraction of 
doubly ionized carbon.
Higher stellar luminosities and lower abundances
give even higher ratios.

These models can be compared with the observed values
for these galaxies, after correcting the H$\alpha$ fluxes
for extinction and 
contributions from diffuse gas.
For NGC 4294 and NGC 4299, we estimate average
extinctions of A$_{\rm V}$ $\sim$ 0.5 over the ISO beam,
using the HI column densities
from \markcite{w88b}Warmels (1988b)
and the standard Galactic relationship between
hydrogen column density and extinction from
\markcite{bsd78}Bohlin, Savage, $\&$ Drake (1978),
assuming
molecular gas is negligible
(i.e., assuming the standard Galactic
I$_{CO}$/N$_{H_2}$
ratio).
We assume $\sim$35$\%$ of the observed
H$\alpha$ emission from these Virgo
galaxies comes from the diffuse ionized gas (warm
ionized medium; WIM), as in
M31, the Magellanic Clouds, and other
nearby spiral galaxies \markcite{wb94}(Walterbos
$\&$ Braun 1994; \markcite{k95}Kennicutt et al. 1995;
\markcite{hwg96}Hoopes, Walterbos, $\&$ Greenawalt 1996).
This calculation indicates that 10 $-$ 20$\%$
of the total observed [C~II] emission from these
galaxies may arise from classical HII regions, if 
the ionizing stars have temperatures between
31,000K and 37,000K, the HII region densities 
are
$\sim$100 cm$^{-3}$, and the abundances
are solar.  For higher stellar temperatures and gas
densities
or lower abundances,
this fraction decreases.
On the other hand, if the extinction is underestimated,
HII regions may contribute a larger portion of
the observed [C~II] flux.

\markcite{r85}Rubin (1985) 
does not model low density
HII regions, so we estimate their contributions via the
relationship
M(HII)/L([C~II]) = [ 0.5 + T$^{1/2}$/(15X$_e$n$_H$) ] 
(3 $\times$ 10$^{-4}$/X$_{C^+}$).
In this formula, M(HII) and L([C~II])
are in solar units, T is in degrees K, and n$_H$ is
in cm$^{-3}$.
This equation was derived from the above relationship
for the WNM, using
the fact that
in ionized gas, excitation is dominated
by collisions with electrons.
Using solar abundances and typical values for the WIM
in the Milky Way
of 8000K, n$_H$ $\sim$ 0.5~cm$^{-3}$,
and X$_e$ $\ge$ 0.75 \markcite{kh87}(Kulkarni $\&$ Heiles 1987),
the ratio 
M(HII)/L([C~II]) 
$\sim$ 15~M$_{\sun}$/L$_{\sun}$.

The mass of diffuse ionized hydrogen in these galaxies can
be estimated from their extinction-corrected H$\alpha$ luminosities.
Using the case B 10,000K relationship
\markcite{o89}(Osterbrock 1989), 
L(H$\alpha$) $\sim$ 0.11 M(HII) n$_H$,
where 
L(H$\alpha$) is in L$_{\sun}$, M(HII) is in M$_{\sun}$,
and n$_H$ is in cm$^{-3}$.
For the WIM, 
where n$_H$ $\sim$ 0.5~cm$^{-3}$,
L(H$\alpha$) $\sim$ 0.55 M(HII).
Assuming 35$\%$ of the observed
H$\alpha$ emission 
comes from the WIM, 
we estimate that the HII mass in the WIM
is $\sim$2.4 $\times$ 10$^7$ M$_{\sun}$
in these galaxies.
The expected [C~II] luminosity
is therefore 1.6 $\times$ 10$^6$~L$_{\sun}$
for the WIM.

We therefore conclude that ionized hydrogen regions contribute
$\le$35$\%$ of the total observed [C~II]
emission from these galaxies, with roughly 15$\%$
originating in the WIM and $\le$20$\%$ in standard HII regions.
These percentages are uncertain because the extinction
is poorly determined, as is the density and the ratio of 
diffuse to dense
HII mass.
Multi-frequency high spatial resolution radio
continuum observations and deep H$\alpha$ imaging
of these galaxies
would be helpful in better determining the contributions
from these components to the observed [C~II] emission.

As noted previously, NGC 4294 and NGC 4299
may have relatively low 
L([N~II])/L([C~II]) 
ratios compared
to the starburst galaxy M82.
Since 
the ionization energy of nitrogen is less than that
of hydrogen
and 
[N~II] is only found in HII regions,
these low 
L([N~II])/L([C~II]) 
ratios suggest
that C$^+$ associated 
with ionized hydrogen may be less important in NGC 4294
and NGC 4299 than in starburst galaxies such as M82.
This result is quite uncertain, however, because of the calibration
uncertainties and the lack
of published [N~II] measurements for the WIM.  
This analysis emphasizes the need for spatially
resolved far-infrared line studies of nearby galaxies, such
as the 
\markcite{m93}Madden et al. (1993) study, to better
distinguish the various components of [C~II] emission.

\section{Conclusions}

We have observed the [C~II] 158 $\mu$m, [N~II] 122 $\mu$m,
and CO (1 $-$ 0) emission from five 
Virgo spiral galaxies with
low blue luminosity.  In three of
the sample galaxies, the observed fluxes are consistent
with emission from solar metallicity PDRs.
In two out of the five galaxies
in the sample, however, we find high
L([C~II])/L(CO) and L(FIR)/L(CO) ratios, 
more typical of dwarf irregular
galaxies than of high mass spirals.  
These enhanced ratios may be caused by low metallicities,
and therefore non-Galactic 
I$_{CO}$/N$_{H_2}$
ratios.  
Alternatively, 
C$^+$ emission from
cold neutral HI clouds rather
than PDRs may dominate the [C~II] emission
in these galaxies, or both HI-dominated
regions and low metallicity PDRs may contribute to the observed [C~II] 
emission.
The WIM and standard HII regions 
contribute less than a third of the observed [C~II] emission.

\acknowledgments

We are grateful to the entire ISO team for all their
hard work on ISO.
We particularly thank the LWS instrument team
and the Vilspa ground support, as well as the 
ISO staff at the Infrared Processing
and Analysis Center (IPAC).
We also thank Jeff Kenney, Steve Lord,
Phil Maloney, and an anonymous referee for helpful comments on the manuscript.
We appreciate the help of the NRAO staff in making
the CO observations, particularly the telescope
operators Lisa Engel, Paul Hart, Mark Metcalf, and Victor Gasho.
This research has made use of the NASA/IPAC Extragalactic
Database (NED) which is operated by the Jet Propulsion Laboratory,
Caltech, under contract with the National Aeronautics and Space
Administration.
A portion of this work was performed while B.J.S.
held a National Research Council Research Associateship
at the Jet Propulsion Laboratory.
This work was supported in part by ISO data analysis funding from the US
National Aeronautics and Space Administration.

\clearpage

{\bf CAPTIONS}

\figcaption[Smith.fig1.ps]{Left column: the final summed CO (1 $-$ 0)
spectra for the central positions in the 
sample galaxies.
Note that these plots all have the same y-axis scale.
Right column: the final [C~II] 158 $\mu$m spectra
for the sample galaxies.  These data are all plotted
with the same y-axis scale. \label{fig1}}

\figcaption[Smith.fig2.ps]{Figure 2. The final [N~II] 122 $\mu$m spectra
for the three observed galaxies. \label{fig2}}

\end{document}